# Teaching an Old Elephant New Tricks

Nicolas Bruno
Microsoft Research
nicolasb@microsoft.com

## ABSTRACT

In recent years, column stores (or C-stores for short) have emerged as a novel approach to deal with read-mostly data warehousing applications. Experimental evidence suggests that, for certain types of queries, the new features of C-stores result in orders of magnitude improvement over traditional relational engines. At the same time, some C-store proponents argue that C-stores are fundamentally different from traditional engines, and therefore their benefits cannot be incorporated into a relational engine short of a complete rewrite. In this paper we challenge this claim and show that many of the benefits of C-stores can indeed be simulated in traditional engines with no changes whatsoever. We then identify some limitations of our "pure-simulation" approach for the case of more complex queries. Finally, we predict that traditional relational engines will eventually leverage most of the benefits of C-stores natively, as is currently happening in other domains such as XML data.

## 1. MOTIVATION

In the last couple of decades, new database applications have emerged with different requirements than those in traditional OLTP scenarios. A prominent example of this trend are data warehouses, which are characterized by read-mostly workloads, snowflake-like schemas, and ad-hoc complex aggregate queries. To address these scenarios, the database industry reacted in different ways.

On one hand, traditional database vendors (e.g., Microsoft, IBM, and Oracle) augmented traditional database systems with new functionality, such as support for more complex execution plans, multi-column index support, and the ability to automatically store, query and maintain materialized views defined over the original data.

On the other hand, new players in the database market devised a different way to store and process read-mostly data. This line of work was pioneered by Sybase IQ [1] in the mid-nineties and subsequently adopted in other systems [7, 15]. The main idea in such column-oriented stores (also called *C-stores*) is to store data column-by-column rather than the traditional *row-by-row* approach used in traditional systems (called *row-stores* in this context). Since queries read only the columns that they truly require, query processing in C-stores becomes more efficient. Additionally, storing data by column results in better compression than what is possible in a row-store. Some compression techniques used in C-stores (such as dictionary or bitmap encoding) can also be applied to row-stores. However, RLE encoding, which replaces a sequence of the same value by a pair (value, count) is a technique that cannot be directly used in a row-store, because wide tuples rarely agree on all attributes. The final ingredient in a C-store is the ability to perform query processing over compressed data as much as possible (see [5] for an in-depth study on C-stores).

C-stores claim to be much more efficient than traditional row-stores. The experimental evaluation in [15] results in C-stores being 164x faster on average than row-stores, and other evaluations [14] report speedups from 30x to 16,200x (!). These impressive results make us wonder whether we could incorporate some of the beneficial features of C-stores in traditional row-stores to obtain a system that performs very well not only in specific data-warehouse vertical, but throughout the spectrum of database applications. Unfortunately, some proponents of C-store architectures claim that their design principles are so different from those in row-stores that they cannot be effectively emulated [6], and moreover that *"it will require widespread and extensive code modifications for row-stores to even approach column-store performance"* [4].

In this paper we challenge this claim by investigating ways to simulate C-stores inside row-stores. In Section 2 we show how to exploit some of the distinguishing characteristics of C-stores inside a row-store *without any engine changes*. Then, in Section 3 we discuss some limitations of this approach and predict how row-stores would eventually incorporate most of the benefits of a C-store without losing the ability to process non data-warehouse workloads.

## Experimental Setting

All our experiments were conducted using an Intel Xeon 3.2GHz CPU with 2GB of RAM and a 250GB 7200RPM SATA hard drive running Windows Server 2003 and Microsoft SQL Server 2005. To validate our results, we use the same data set and workload proposed in the original C-store paper [15]. Specifically, we used a TPC-H database with scale factor ten and the seven queries[1] of Figure 1. Although additional data sets and workloads have been used in subsequent papers, the one in [15] is a representative micro-benchmark particularly well suited for C-stores and therefore a good "stress test" for our approach. Following [15], we assume that the following schema is used in the C-store:

```
D1:   (lineitem | l_shipdate, l_suppkey)
D2:   (lineitem ⋈ orders | o_orderdate, l_suppkey)
D4:   (lineitem ⋈ orders ⋈ customer | l_returnflag)
```

---
[1]Reference [15] does not specify the D values for queries with inequalities on date columns (i.e., $Q_1$, $Q_3$, $Q_4$, and $Q_6$). Therefore, in our experiments we used values of D that resulted in a wide range of selectivity values.

- $Q_1$ (count of items shipped each day after D):
  ```
  SELECT l_shipdate, COUNT (*)
  FROM lineitem
  WHERE l_shipdate > D
  GROUP BY l_shipdate
  ```
- $Q_2$ (count of items shipped for each supplier on day D):
  ```
  SELECT l_suppkey, COUNT (*)
  FROM lineitem
  WHERE l_shipdate = D
  GROUP BY l_suppkey
  ```
- $Q_3$ (count of items shipped for each supplier after day D):
  ```
  SELECT l_suppkey, COUNT (*)
  FROM lineitem
  WHERE l_shipdate > D
  GROUP BY l_suppkey
  ```
- $Q_4$ (latest shipdate of all items ordered after each day D):
  ```
  SELECT o_orderdate, MAX (l_shipdate)
  FROM lineitem, orders
  WHERE l_orderkey=o_orderkey AND o_orderdate>D
  GROUP BY o_orderdate
  ```
- $Q_5$ (for each supplier, latest shipdate of an item from an order that was made on day D):
  ```
  SELECT l_suppkey, MAX (l_shipdate)
  FROM lineitem, orders
  WHERE l_orderkey=o_orderkey AND o_orderdate = D
  GROUP BY l_suppkey
  ```
- $Q_6$ (for each supplier, latest ship date of an item from an order that was made after day D):
  ```
  SELECT l_suppkey, MAX (l_shipdate)
  FROM lineitem, orders
  WHERE l_orderkey=o_orderkey AND o_orderdate > D
  GROUP BY l_suppkey
  ```
- $Q_7$ (Nations for customers (along with lost revenue) for parts that they returned):
  ```
  SELECT c_nationkey, SUM(l_extendedprice)
  FROM lineitem, orders, customers
  WHERE l_orderkey=o_orderkey AND o_custkey=c_custkey
    AND l_returnflag='R'
  GROUP BY c_nationkey
  ```

**Figure 1: Queries used in the experimental evaluation.**

where $D_i$ = (expression | sortCols) means that we individually materialize all columns in expression after sorting it by sortCols.

In addition to our proposed strategies, we evaluate two baseline query processing techniques:

**Row:** Corresponds to the traditional query processing by a row-store for which only primary indexes have been materialized.

**ColOpt:** Corresponds to a (loose) lower bound on any C-store implementation. We achieve this lower bound by manually calculating how many (compressed) pages in disk need to be read by any C-store execution plan, and measuring the time taken to just read the input data. In other words, we do not consider any filtering, grouping or aggregation over the input data, and thus this strategy represents the absolute minimum time taken by any C-store implementation. We decided to use this baseline to avoid directly comparing different systems written and optimized by different groups of people.

Figure 2 shows the execution times of both Row and ColOpt for the queries in Figure 1 (ignore the additional bars for now)[2]. The immediate conclusion from the figure is that column stores indeed have the *potential* to result in very large speedups with respect to plain row-stores, as illustrated more concisely in the table below:

|  | $Q_1$ | $Q_2$ | $Q_3$ | $Q_4$ | $Q_5$ | $Q_6$ | $Q_7$ |
| --- | --- | --- | --- | --- | --- | --- | --- |
| Speedup | 26,191x | 4,602x | 59x | 35x | 2,586x | 37x | 113x |

We next explore how we can improve the performance of Row towards that of ColOpt without any changes to existing systems.

## 2. SIMULATING C-STORES

There has been previous work on simulating a C-store inside a row-store (e.g., see [3, 4, 6]). The idea is to replace each table with either vertical partitions or single-column non-clustered indexes. These references show that both approaches fail to deliver good performance (in fact, in general they perform even worse than plain row-stores). The reason (not unexpected in hindsight) is that single column indexes or partitions cannot be stored or processed in compressed form due to mandatory extra tuple information (e.g., rids must be present in secondary indexes or vertical partitions), and also that tuples in different partitions or indexes are sorted in different ways, which result in many index seeks when trying to combine multiple column values.

### 2.1 Varying the Physical Design

An advantage of C-stores is that they support some sort of "pre-computed" representation of each column via RLE compression. In fact, the values in a column are first sorted[3], and then each sequence of $k$ instances of the same value $v$ is replaced by the pair $(v, k)$. This mapping considerably reduces the space required to store columns (especially those earlier in the global ordering) and also speeds up query processing of filters and aggregates due to some information being already pre-aggregated at the storage layer.

Interestingly enough, row-stores have invested considerably on sophisticated mechanisms to store and process pre-computed information, commonly denoted materialized views. Materialized views not only store information in aggregated form, but can also be used to answer queries that do not match exactly the view definition, and are automatically updated. Materialized view languages are rich enough that many of the queries in Figure 1 can be directly pre-materialized using views. For instance, $Q_7$ in Figure 1 can be converted into a materialized view, and then answering $Q_7$ would just entail reading the answer from disk. This approach would not work if we change parameter values (e.g., changing l_returnflag='R' to l_returnflag='A' would prevent the view from being matched). We therefore generalize the view definitions to match any instance of queries $Q_1$ to $Q_7$ when arbitrarily varying their parameter values, as illustrated below for queries $Q_2$, $Q_3$ and $Q_7$:

```
MV2,3 =   SELECT l_shipdate, l_suppkey, COUNT(*)
          FROM lineitem
          GROUP BY l_shipdate, l_suppkey

MV7 =     SELECT c_nationkey, l_returnflag, SUM(l_extendedprice)
          FROM lineitem, orders, customers
          WHERE l_orderkey=o_orderkey AND o_custkey=c_custkey
          GROUP BY l_returnflag, c_nationkey
```

Figure 2 contrasts the execution time of the strategies that implement materialized views natively in SQL Server 2005 (Row(MV) in the figure) and the loose lower bound of any C-store implementa-

---
[2] Due to equality predicates in columns D and l_returnflag, execution times for queries $Q_2$, $Q_5$, and $Q_7$ do not depend on parameter values. We therefore show results for a single value of the parameters.

[3] Note that column $c$ is not necessarily sorted in $c$ order, but follows instead the global ordering defined by a DBA. Column correlation, however, still produces clusters of the same value in $c$, which is therefore compressed.

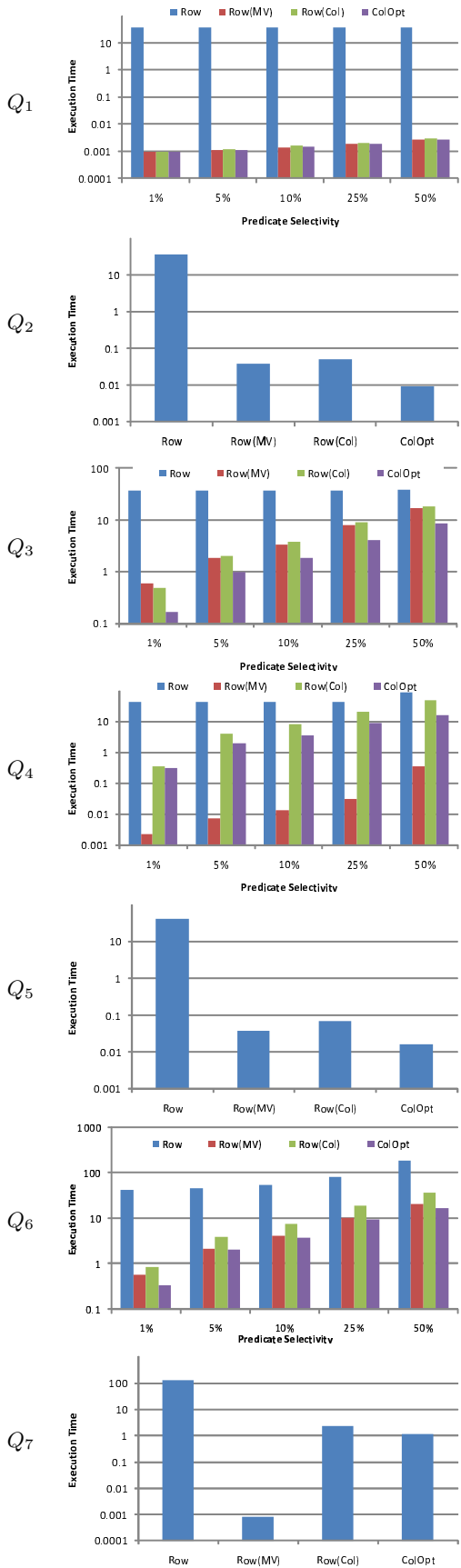

**Figure 2: Results of the experimental evaluation.**

tion (`ColOpt` in the figure). The table below summarizes the average relative performance of `Row(MV)` compared to `ColOpt`:

| | $Q_1$ | $Q_2$ | $Q_3$ | $Q_4$ | $Q_5$ | $Q_6$ | $Q_7$ |
|---|---|---|---|---|---|---|---|
| Row(MV) | = | 4x↑ | 2x↑ | 250x↓ | 2.5x↑ | 1.2x↑ | 1,400x↓ |

Using materialized views for $Q_2$ results in a plan that is 4x slower than the lower bound for C-stores, and for $Q_7$ in a plan that is 1,400 times better than the best possible C-store implementation. These results are interesting, since, as we discussed earlier, the time for `ColOpt` only considers reading the compressed input values, but does not take into account and subsequent query processing. For instance, we measured $Q_2$ for the `Row(MV)` case and found that roughly 40% of the execution time is spent grouping and aggregating results (in some form or another, that overhead must also be present in any implementation of C-stores, bringing the already modest speedup further down). In conclusion, while some queries *could be* executed at most 2-4 times more efficiently in a C-store implementation, others can be hundreds or even thousands of times more efficient by using materialized views.

While the performance of the workload using materialized views is impressive (and could be made even more efficient by using the compressed representation of row-stores proposed in [10]), the main drawback is generality. While materialized views can answer queries that are slightly different from the view definition (e.g., changing a constant value for another) they would not match other common modifications. This might not be an issue in scenarios that contain mostly reporting queries (and it should be, in fact, the right approach), but can become a significant problem for application that issue significant number of ad-hoc queries. We next explore a different approach for simulating C-store benefits inside a row-store without modification to traditional engines.

## 2.2 Varying the Logical Design

So far we discussed two extreme physical designs. On one hand, single-column indexes are flexible for varying workloads, but generally result in inefficient executions [6]. On the other hand, materialized views are extremely efficient but a bit narrow in scope. We now present a technique that is based on changing the *logical database design*, requires no modification to current query engines, and results in efficient executions (close to those of C-stores) without suffering from the specificity of materialized views.

### 2.2.1 Logical Database Design using C-Tables

The main idea of our approach is to extend the vertical partition approach in [6] to explicitly enable the RLE encoding of tuple values. Concretely, consider a table $T$ with columns $a$, $b$, and $c$, as shown in Figure 3(a). Also, suppose that we want to simulate the

| (virtual) id | a | b | c |
|---|---|---|---|
| 1 | 1 | 1 | 1 |
| 2 | 1 | 1 | 4 |
| 3 | 1 | 2 | 4 |
| 4 | 1 | 2 | 5 |
| 5 | 1 | 2 | 5 |
| 6 | 2 | 1 | 1 |
| 7 | 2 | 1 | 1 |
| 8 | 2 | 3 | 1 |
| 9 | 2 | 3 | 2 |
| 10 | 2 | 3 | 2 |
| 11 | 2 | 3 | 3 |
| 12 | 2 | 3 | 4 |

| $T_a$ | f | v | c |
|---|---|---|---|
| | 1 | 1 | 5 |
| | 6 | 2 | 7 |

| $T_b$ | f | v | c |
|---|---|---|---|
| | 1 | 1 | 2 |
| | 3 | 2 | 3 |
| | 6 | 1 | 2 |
| | 8 | 3 | 5 |

| $T_c$ | f | v |
|---|---|---|
| | 1 | 1 |
| | 2 | 4 |
| | 3 | 4 |
| | 4 | 5 |
| | ... | ... |

(a) Original Table.     (b) Logical Representation.

**Figure 3: Logical database design for row-stores.**

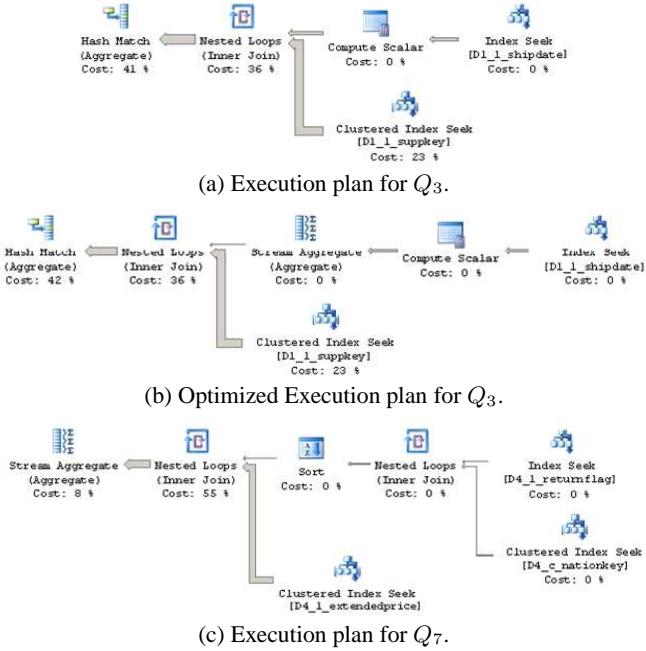

(a) Execution plan for $Q_3$.

(b) Optimized Execution plan for $Q_3$.

(c) Execution plan for $Q_7$.

**Figure 4: Execution plans using c-tables.**

C-store schema $(T|a, b, c)^4$ (i.e., we sort by $a$, then $b$, and finally $c$). We then proceed (conceptually) as follows. First, we sort the table according to the sort columns in the schema and associate to each tuple a virtual column $id$ that represents the position of the tuple in the resulting ordering (see Figure 3(a)). Then, we generate one table $T_x$ per column $x$ in the schema (we call these *c-tables*). To do so, we "group" each sequence of the same value for $c$ in the sorted table that additionally agree with all the previous sort columns. We then represent each of these groups in the original table by a tuple $(f, v, c)$ in the c-table $T_x$, where $v$ is the repeated value in the group, $f$ is the minimum $id$ value in the original table for the elements in the group, and $c$ is the group size. Figure 3(b) shows an example of this mapping for the table in Figure 3(a). Note that some columns (especially those deep in the sort order), will have most of the $c$ values equal to one. If this happens for column, say, $x$, the alternative representation that simply projects columns $id$ and $x$ from the original table might be smaller and we use it instead (e.g., see $T_C$ in Figure 3). Once the tables have been populated, we create a clustered index on column $f$ and a secondary covering index with leading column $v$ (as we show later, the ability to have multiple indexes in c-tables enables additional query execution plans).

The meaning of a tuple $(f, v, c)$ in c-table $T_x$ is that, on the original table, all tuples from position $f$ to position $f + c - 1$ have value $v$. An interesting property of this logical representation is that for any pair of tuples $t_1$ and $t_2$, possibly on different c-tables, the ranges $[f_1, f_1 + c_1 - 1]$ and $[f_2, f_2 + c_2 - 1]$ do not partially overlap (i.e., they are either disjoint or the one associated with the column deeper in the sort is included in the other). This fact allows us to exploit specific rewritings to combine information from different c-tables to answer queries.

### 2.2.2 Query Rewriting

Query rewriting for c-tables is an extension of the traditional mechanisms to answer queries using vertical partitions, in which we leverage the compressed representation of columns. The main idea is to join different c-tables with "band" joins and take advantage of the compressed representation during query processing. Consider query $Q_3$ in Figure 1 and the logical design derived from schema D1:(lineitem | l_shipdate, l_suppkey). We then rewrite $Q_3$ as follows:

```
SELECT T1.v, SUM(T1.c)
from D1_l_suppkey T1, D1_l_shipdate T0
WHERE T0.v>D
  AND T1.f between T0.f and T0.f+T0.c-1
GROUP BY T1.v
```

where (i) we join c-tables D1_l_shipdate T0 and D1_l_suppkey T1 using the predicate T0.f $\leq$ T1.f $<$ T0.f+T0.c-1, which exploits the property of c-table ranges described above, and conceptually returns compressed 4-tuples (T1.v, T0.v, T0.f, T0.c), and (ii) we replace the count(*) aggregate value with sum(T1.c), which effectively performs the aggregation over compressed data. The execution plan for this query is shown in Figure 4(a). It first performs an index seek on the secondary index for the c-table corresponding to l_shipdate and returns all tuples that satisfy v > D. For each resulting tuple (f,v,c), it seeks for all tuples in the c-table corresponding to l_suppkey for the corresponding matches, which are then grouped and aggregated using a hash-based operator.

[4]For simplicity, in this work we assume that all the columns in $T$ appear in the sort columns in the schema. We will later comment on this assumption.

### 2.2.3 Optimizations

The logical schema described above, along with the available indexes and advanced query processing strategies of the query engine allow different *mechanizable* query optimizations:

**Query-specific rewriting rules:** Consider again query $Q_3$ and the plan in Figure 4(a). Note that column l_shipdate is used only to restrict tuples, but not used in the final result, which only groups by l_suppkey. Additionally, note that tuples satisfying l_shipdate > D are clustered in table D1_l_shipdate and therefore the ranges [f, f+c-1] are all consecutive. We can rewrite $Q_3$ as follows:

```
SELECT T1.v, SUM(T1.c)
FROM (SELECT MIN(T0.f) AS xMIN,
             MAX(T0.f+T0.c-1) AS xMAX
      FROM D1_l_shipdate T0 WHERE T0.v>D) T0Agg,
     d1_l_suppkey T1
WHERE T1.f BETWEEN T0Agg.xMin AND T0Agg.xMax
GROUP BY T1.v
```

This query results in the execution plan of Figure 4(b), which has much fewer context switches since there is a single tuple in the outer side of the nested loop join.

**Rich query-processing engine:** The complex query engine in the row-store makes possible non-obvious, cost-based optimizations. Consider $Q_7$, rewritten as follows:

```
SELECT T1.v, SUM(T2.c*T2.v)
FROM d4_l_returnflag T0,
     d4_c_nationkey T1,
     d4_l_extendedprice T2
WHERE T0.v='R'
  AND T2.f BETWEEN T1.f AND T1.f+T1.c-1
  AND T1.f BETWEEN T0.f AND T0.f+T0.c-1
GROUP BY T1.v
```

The execution plan is shown in Figure 4(c). Note that after the first join that puts together columns o_orderdate and l_suppkey, the query processor introduces an intermediate sort operator, which produces tuples sorted by T1.v values that are later aggregated using a stream-based operator.

**Additional index-based strategies:** Multiple indexes on c-tables (e.g., covering indexes on v values) enable additional strategies. Consider schema (T | a,b,c,d) and the query below:

```
SELECT a, b, c, d
FROM T
WHERE c=10 AND d=20
```

Note that the predicates are over columns deep in the sort order. Therefore, C-store implementations would have to either scan the full c and d columns (perhaps using late materialization), or perform seeks over c (and then d) *for each* combination of (a,b) values, which could be even more expensive if there are many distinct values. Instead, we can easily evaluate the query by (i) seeking *both* c-tables independently using indexes on v values and "intersecting" partial results, and (ii) obtaining the remaining columns as usual. This strategy can be more efficient than any C-store alternative.

### 2.2.4 Experimental Results

We implemented our approach by materializing c-tables and executing the rewritten queries from Figure 1. Figure 2 shows the results of our techniques (denoted Row(Col) in the figure). The table below summarizes the average slowdown of our techniques relative to the loose lower bound of any possible C-store implementation:

|  | $Q_1$ | $Q_2$ | $Q_3$ | $Q_4$ | $Q_5$ | $Q_6$ | $Q_7$ |
|---|---|---|---|---|---|---|---|
| Row(Col) | 1.1x | 5.6x | 2.3x | 2.2x | 4.2x | 2.1x | 2.0x |

The performance of the workload is on average 2.7x slower than the lower bound for any C-store implementation. Considering that (i) such implementation would have to additionally perform filters, grouping and aggregation, and (ii) row-store execution plans can coexist with our strategies (it is just a different logical design), we believe that our approach enables both performance and flexibility rivaling those of C-stores in an plain, *unmodified* row-store.

## 3. A PEEK INTO THE FUTURE

Although the results in the previous section are very encouraging, there are opportunities to further improve performance. At the same time, the ideas discussed below can help mitigate certain limitations that appear when translating and fine-tuning more complex queries with our self-imposed restriction of making no changes whatsoever to a traditional engine.

**Storage layer:** Although columns are effectively stored using RLE encoding, we still use additional overhead per (compressed) tuple. Our row-store system uses 9 bytes of overhead per tuple, which can effectively double the amount of space required to store data in a native C-store. Small changes in page layout would certainly improve this problem [10], especially because c-tables are clustered by increasing and dense f values, which can be effectively delta-compressed.

**Column concatenation:** We forced every schema to have deep sort orders, in which all columns participated in the sort. In some scenarios, having shorter sort sequences allow better compression of the remaining columns. In such cases, we need to use more complex band predicates to join columns together. While this is possible, the optimizer and query engine do not recognize certain high level properties of the data, which prevents efficient query processing strategies. An alternative approach is to create user-defined operators using C# table-valued functions in our database system. These extensions would take two streams and "concatenate" them into one, similarly to what C-stores do. We implemented such approaches but they are not particularly efficient (they are outside the server, the logic is quasi-interpreted and performance is not high enough). Changes in the optimizer and execution engine would mitigate this issue.

**Query hints:** We had to sometimes hint the query optimizer to pick a different plan than the default one. For instance, sometimes merge joins are picked over index nested loop joins because the optimizer assumes that each index seek from a tuple for the outer relation would incur random I/Os. However, the properties of our data is such that all requests are strictly sorted, which results in much lower execution times. Additionally, sometimes cardinality estimates are not as accurate as they could be by exploiting the semantics of our data representation. In general, the optimizer lacks domain-specific information to make better choices. Adding such logic would remove the need to specify query hints.

**Software development:** Our techniques require a careful rewriting of the original queries into alternatives that are much more difficult to understand and maintain. However, all the translation mechanisms are mechanical, and can be easily incorporated into a middleware framework such as LINQ [2], which enables such mappings as first class citizens. Application developers would then issue queries without considering whether the underlying representation (and execution plans) follow the traditional row-store architecture, the strategies discussed in the previous section, or even a hybrid of both.

In conclusion, although for all the queries in Figure 1 we get most of the benefits of C-stores, there are still obstacles that prevent a full simulation within a row-store. At first sight it seems that several components in a traditional DBMS have to be adapted in some form or another for this to be possible. Is this task monumental, as some C-store proponents argue, or requires just evolutionary changes similar to those that already happened in similar contexts?

### 3.1 Lessons From The (Near) Past

One way to concisely summarize the limitations of our approach is that traditional engines cannot leverage the very specific semantics of c-tables. Rather than identifying and exploiting their special characteristics at both the storage and query processing layers, the system considers them as regular tables. We next review two series of events from the recent past that share several characteristics with our methodology and are therefore very relevant to our discussion.

#### 3.1.1 XML Query Processing

In the last decade, XML emerged as a de facto standard for information representation and exchange over the internet. XML data and query models are rather different from those in the relational model, and it was argued that native engines were the only approach to effectively query XML data. Over the years, relational engines went through two phases to incorporate XML support.

**Mid-Tier Approach:** Several pieces of work adopted a mid-tier approach that consisted of mapping an XML schema into the relational schema (e.g., see [11]). XML data is shredded into relational form using a schema-driven approach and queried using XPath, which is internally translated into SQL queries. While this approach was successful, the question of extending query rewrites for more complex XML fragments remained open, and some complex scenarios stretched the capabilities of relational systems, which had no domain-specific knowledge about the shredded schema.

**Native Support:** Driven by both performance and expressiveness requirements, DBMS vendors started pushing XML functionality inside the query engine (e.g., see [13]). As an example, rather than building a new XML-only system, SQL Server 2005 deeply integrated the XML capabilities into the existing framework, which provides general services such as backup, restore, replication, and concurrency control. While this was not an easy task, the implementation was surprisingly componentized, and the resulting system is able to both leverage the relational storage and query infrastructure, and support XML query processing in a seamlessly integrated way (i.e., XML, purely relational, and hybrid queries coexist in the same system and build upon the same technology).

### 3.1.2 Spatial Query Processing

Location-aware devices and services are increasingly becoming commodity items. Many scientific applications rely on spatial operations over real-world objects. While each spatial application seems to have some unique characteristic, over the years there has been consensus on a series of primitives to perform spatial query processing [12]. Similarly to the XML scenario, relational engines went through phases to incorporate spatial support.

**Spatial Library:** Reference [9] describes a library that is built on top of a traditional database system and can perform spatial query processing. The idea is to store spatial data in specialized but plain SQL tables, and rely on SQL functions and stored procedures to provide primitive spatial operators. Pushing the logic entirely into SQL allows the query optimizer to do a very efficient job at filtering relevant objects.

**Native Support:** Recently, database systems started incorporating native spatial support to the query engines (e.g., see [8]). For instance, SQL Server 2008 added spatial data support to manage location-aware data. In particular it introduces two new built-in types to represent planar and geodetic vector data. An adaptive, multi-level grid based spatial index provides efficient processing, and it is built on existing B+-Tree infrastructure and integrated into the query optimizer. As with XML, spatial data coexists with traditional relational tuples, leverages the same basic infrastructure and all the peripheral utilities and features of the DBMS.

### 3.1.3 What About C-stores?

A pattern that emerges from the previous examples is as follows. First, new classes of applications introduce novel requirements. This is followed by specialized query engines that are domain specific, very efficient, but at the same time, understandably narrow in scope. A little later, some of the specialized functionality is simulated as a thin layer on top of traditional relational systems. Finally, the functionality is packaged into a new data type, index, query processing technique or a combination of these, and supported natively inside a relational system. Native support inside a generic DBMS (at least in initial releases) is usually less efficient and possibly more restrictive than a specialized engine. However, when the new functionality becomes stable, performance is usually comparable, and the approach additionally results in significant side benefits: (i) all the development, deployment, and testing tools, among others, are transparently supported for the new functionality, (ii) maintaining one complex system is easier than maintaining multiple (also complex!) systems, and (iii) supporting applications with hybrid requirements is easier without additional data migration support.

In this paper we claim that C-stores are going through the "thin-layer simulation" phase, in which many of its benefits are captured with no changes to the kernel of a traditional row-store. Of course, we are aware of the limitations of our approach (which we believe are analogous to those in the XML or spatial scenarios described above). Can row-stores go the extra mile and natively incorporate C-store functionality? In principle, we do not see any conceptual obstacle. We predict that this will not be a straightforward task, but at the same time nothing out of the ordinary and similar to what happened multiple times in the recent past.

## 4. CONCLUSIONS

We claim that row-stores have been considered and discarded too quickly for read-mostly, data warehousing scenarios. We think several decades of active research and development have resulted in relational database systems that are robust and extensible, and that can cope with new, unexpected scenarios in a reasonably agile manner. Specifically, we show that relatively simple mappings at the logical database level can result in execution strategies in row-stores that rival those of C-stores. We also predicted that natively understanding C-stores can further close the gap, as it happened before for XML or spatial data. Whether row-stores would be able to incorporate all the benefits of C-stores is something yet to be seen, but we remain cautiously optimistic about this outcome.

## Acknowledgments

We thank Ravi Ramamurthy, Vivek Narasayya and Paul Larson for valuable feedback to an earlier version of this document.